\documentclass[reqno, 11pt]{amsart}
\usepackage{url}

\usepackage{color}
\usepackage{bbm}
\usepackage{pdfsync}
\usepackage{enumerate}
\usepackage{graphicx}
\usepackage{multirow}
\usepackage{amssymb, amsmath}

\numberwithin{equation}{section}

\newcommand{\eqmean}[2]{\langle #1 \rangle^{(\text{eq},#2)}}

\newcommand{\mean}[1]{\langle #1 \rangle}

\newcommand{\muss}{\mu_{\text{s,s}}}
\newcommand{\musv}{\mu_{\text{s,v}}}
\newcommand{\nessvf}[1]{\langle #1 \rangle_{\text{s,v}}}
\newcommand{\nesssc}[1]{\langle #1 \rangle_{\text{s,s}}}

\newcommand\bq{{\mathbf q}}
\newcommand\bp{{\mathbf p}}

\newcommand\ZZ{{\mathbb Z}}

\newcommand\ve{\varepsilon}

\newcommand{\mc}[1]{{\mathcal #1}}

\title[Harmonic Chains with Bulk Noises]{Nonequilibrium Stationary States of Harmonic Chains with Bulk Noises}

\date{\today}

\keywords{Non-equilibrium systems, long-range correlations.}

\begin{document}

\begin{abstract}
We consider a chain composed of $N$ coupled harmonic oscillators in contact
with heat baths at temperature $T_\ell$ and $T_r$ at sites $1$ and $N$
respectively. The oscillators are also subjected to non-momentum conserving
bulk stochastic noises. These  make the heat conductivity  satisfy Fourier's law.  Here we describe some new results about the hydrodynamical equations for typical macroscopic energy and displacement profiles, as well as their fluctuations and large deviations, in two simple models of this type.
\end{abstract}

\author{C.~Bernardin}
\address{C. Bernardin\\
Universit\'e de Lyon and CNRS, UMPA, UMR-CNRS 5669, ENS-Lyon,\\
46, all\'ee d'Italie, \\
69364 Lyon Cedex 07 \\
France
}
\email{Cedric.Bernardin@umpa.ens-lyon.fr}

\author{V.~Kannan}
\address{Venkateshan Kannan\\ Department of Mathematics and Physics\\ Rutgers University\\
Piscataway, NJ\\ USA}
\email{kven@physics.rutgers.edu}

\author{J.~L.~Lebowitz}
\address{Joel L. Lebowitz\\ Department of Mathematics and Physics\\ Rutgers University\\
Piscataway, NJ\\ USA}
\email{lebowitz@math.rutgers.edu}

  \author{J.~Lukkarinen}
\address{Jani Lukkarinen\\
Department of Mathematics and Statistics\\
University of Helsinki\\
P.O.~Box 68\\
00014~Helsingin yliopisto\\
Finland
}
\email{jani.lukkarinen@helsinki.fi}

\maketitle

\section{Introduction}
\label{sec:models}
The time evolution and the nature of the resulting non-equilibrium stationary states (NESS) of systems in contact, at their boundaries, with thermal reservoirs at different temperatures remains a challenging problem.  The only such systems with bulk Hamiltonian dynamics for which NESS are known are harmonic crystals \cite{RLL,Nak}. These are described by specifying the positions and momenta of the oscillators, $q_j$ and $p_j$, $j=1,2, \ldots, N$. The Hamiltonian for a $D=1$ chain (with obvious extension to $D>1$) is given by
\begin{equation}
\label{eq:energy}
{\mc H}{ \bf (q,p)} =\sum_{j=0}^{N+1} \left[ \cfrac{p_j^2}{2}\;  + \;  \nu^2  \cfrac{ q_j^2}{2} \; +\sum_{\substack{|i-j|=1}}\cfrac{(q_{j}-q_i)^2}{4} \right] = \sum_{j=0}^{N+1} {\mc E}_j  
\end{equation}
where we have set the mass of each particle and the nearest neighbor coupling equal to $1$. We impose boundary conditions $q_0=q_{N+1}=0$, $p_0=p_{N+1}=0$. The strength of the pinning potential is regulated by the parameter $\nu \ge 0$. This system is put in contact with thermal reservoirs at temperatures $T_\ell$ at site $1$ and $T_r$ at site $N$ via the Ornstein-Uhlenbeck (Langevin) process.

In the absence of any stochastic bulk noise the heat flux in the non-equilibrium stationary state is independent of  $N$, corresponding to the heat conductivity diverging linearly in $N$ \cite{RLL}. To imitate the effects of interactions between phonons and to produce NESS with normal transport one may add bulk noise to the system \cite{BLL04}--\cite{Ber0}. The new NESS will, we expect, exhibit certain universal features also present in realistic anharmonic systems. It is the purpose of this note to summarize some new results about such systems. A more  detailed analysis will be given elsewhere \cite{BKLL}.

 We shall now consider two ways of adding noise to the system. The first case we consider is the so-called self-consistent model \cite{BLL04}. Each site $j \in \{1,\ldots,N\}$ is connected to a Langevin reservoir at temperature $T_j >0$. 
The temperatures of the reservoirs at the boundary sites are fixed by the conditions  $T_1=T_\ell, \; T_N =T_r$, while the temperatures of the interior site reservoirs are determined self-consistently by requiring that in the NESS there is no net flux of energy between the system and the interior reservoirs.

The second model we consider is one in which we add in the bulk  flip dynamics which consists of reversing the velocity of each particle at random independent Poissonian times.  These flips are energy but not momentum conserving and we call the model the velocity flip model \cite{DKL}.

Let $\mu ({\bf q},{\bf p}, t)$ be the probability density of our system at time $t$. The time evolution of $\mu$ is described by the Fokker-Planck equation 
\begin{eqnarray}
\frac{\partial \mu}{\partial t} &+& \sum_{i=1}^{N} \left[ p_i \frac{\partial \mu}{\partial q_i} -
 \{ (2q_i - q_{i-1} -q_{i+1}) + \nu^2 q_i\} \frac{\partial \mu}{\partial p_i} \right] \nonumber \\   &=& \sum_{\alpha=1,N} \gamma \frac{\partial }{\partial p_{\alpha}}\left[ p_{\alpha} \mu + T_{\alpha} \frac{\partial \mu}{\partial p_{\alpha}} \right] + \sum_{k} B_{k,b} \mu  
\label{eq:FP}
\end{eqnarray}  
where $B_{k,b}$ represents the bulk noise with $b={\rm v}$ for the velocity-flip and $b={\rm s}$ for the self-consistent case: 
 \begin{eqnarray*}
 B_{k,{\rm v}}  \mu &=& \cfrac{\gamma}{2} \Bigl[ \mu (q_1, \ldots, q_N,p_1,\ldots, -p_{k} \ldots, p_N) \\ && \quad - \mu (q_1,\ldots, q_N, p_1, \ldots, p_{k},\ldots p_N) \Bigr]\, ,  \nonumber \\ 
 B_{k,{\rm s}}  \mu &=&   \gamma \frac{\partial }{\partial p_{k}}\left[ p_{k} \mu + T_{k} \frac{\partial \mu}{\partial p_{k}} \right],  \quad {k=2,\ldots,N-1}.
 \end{eqnarray*}
The NESS of the two models, denoted by $\musv$ and $\muss$ respectively, are solutions of (\ref{eq:FP}) with, for the self-consistent model, $T_k$,  $k=2,\ldots, N{-}1$, determined by the  self-consistency condition 
$\muss (p_k^2) =  \nesssc{p_k^2} = T_k$.  

If the temperatures $T_\ell$ and $T_r$ are equal to a common value $T$, the steady state of the self-consistent chain and the steady state of the velocity flip model are both equal to the Gibbs state with temperature $T$ that we denote by $\eqmean{\cdot}{T}$. This is a Gaussian measure with covariance $C_{{\rm{eq}}} (T) = T C_{{\rm {eq}}} (1)$. For $T_\ell \ne T_r$, the NESS of the two models are different.  However,
it is easy to see that both stationary states are centered, and by the results derived in \cite{DKL}, we know that the two point correlation functions of both models coincide, when the rate of flipping and the coupling to the internal reservoirs are the same. Nevertheless, the two measures are very different: $\muss$ is Gaussian but $\musv$ is a non-trivial mixture of Gaussian states. This has a simple non-trivial consequence:
If $f(\bq, \bp)$ is a linear  function of the form
\begin{equation*}
f(\bq,\bp)=\sum_{i=1}^N  \left(a_i q_i + b_i p_i\right) \, ,
\end{equation*}
where $(a_1,\ldots,a_N, b_1, \ldots,b_N)$ are arbitrary real numbers, then
\begin{equation*}
\nessvf{f^4} \ge 3 \left[ \nessvf{f^2} \right]^2 = \nesssc{f^4}\, .
\end{equation*}
This follows from the fact that for a centered Gaussian measure   $ \langle f^4 \rangle  =3 \left[ \langle f^2 \rangle \right]^2$, and $\nessvf{\cdot}$ is a superposition of such measures.

Both models satisfy Fourier's law with the same value of the conductivity. Indeed, since the microscopic energy current across the bond $(k-1,k)$ 
\begin{equation*}
j_{k}^e = -\cfrac{1}{2} (p_k +p_{k-1}) (q_{k} -q_{k-1} ), \qquad k=2, \ldots, N-1 \, ,
\end{equation*}
is a quadratic function of the momenta and positions, its averages over $\nesssc{\cdot}$ and $\nessvf{\cdot}$ are equal. The macroscopic current $J=  \lim_{N \rightarrow \infty}  \left[ \kappa_N (T_{\ell} -T_r)/N \right]$, with $\kappa = \lim_{N \rightarrow \infty} \kappa_N$, is given according to  \cite{BLL04} by 
\begin{equation}
\label{eq:K}
\kappa =\cfrac{1/\gamma}{2+\nu^2 + \sqrt{\nu^2 (\nu^2 +4)}}.
\end{equation}

The outline of the rest of the paper is as follows:
In Section \ref{sec:hl}  we present the macroscopic equations for the density profiles of the bulk conserved quantities in the velocity-flip model. In Section \ref{sec:cor} we show that long-range energy correlations are present in the velocity-flip model but not in the self-consistent one. Section \ref{sec:num} contains numerical simulations of these long-range correlations.

\section{Hydrodynamical scaling limit of the velocity flip model}
\label{sec:hl}

We have to distinguish two cases according to whether $\nu=0$ (unpinned) or $\nu>0$ (pinned). The unpinned case is similar to that investigated by one of the authors in \cite{Ber}.

\subsection{The unpinned chain}
\label{subsec:uc}

When $\nu=0$ the {\textit{bulk}} dynamics conserves two quantities. The first
one is the energy ${\mc H}$. The second one is the deformation, $\sum_{x} r_x$
of the lattice, with $r_x = q_{x+1} - q_x$, $x=0,1, \ldots, N$.  This second conservation law has to be taken into account in the hydrodynamic analysis. 

The energy at a site $x \in \{1, \ldots, N\}$ is now given by
\begin{align}\label{eq:defunpex}
{\mc E}_x = \cfrac{p_x^2}{2} +\cfrac{r_x^2}{4} + \cfrac{r_{x-1}^2}{4}
\end{align}
and ${\mc E}_0 = \frac{1}{4} r_0^2$, ${\mc E}_{N+1} = \frac{1}{4} r_{N}^2$.
To establish the hydrodynamic limits corresponding to the two conservation laws, we look at  the process with time scaled by $N^2$ and space scaled by $N$, i.e., in the diffusive scale \cite{Ber0}. Assume that initially the process is started with a Gibbs local equilibrium measure ${\hat \mu}$ associated with a macroscopic deformation profile $u_0 (q)$ and a macroscopic energy profile ${\varepsilon}_0 (q)$:
\begin{align}\label{eq:locG}
{\hat \mu}  
=
\frac{1}{Z} \prod_{x=0}^{N+1} \exp \left\{ -\beta_0 (x/N) ({\mc E}_x - \tau_{0} (x/N) r_x ) \right\}\, ,
\end{align}
where $T_0 = \beta_0^{-1}$ and $\tau_0$ are the temperature and tension profiles corresponding to the given energy and deformation profiles assumed to be continuous. Then
we have for any macroscopic point $q\in [0,1]$
\begin{equation}
\lim_{N \to \infty} \langle r_{[Nq]} (0) \rangle = u_0(q), \quad  \lim_{N \to \infty} \langle {\mc E}_{[Nq]} (0) \rangle = {\varepsilon}_0 (q) \, ,
\end{equation}
where $[y]$ is the integer part of $y$ and the averages are w.r.t.\ $\hat{\mu}$.

The question then is: what happens at any later (macroscopic) time $t$? It is shown in \cite{BKLL} that
\begin{equation}
\lim_{N \to \infty} \langle r_{[Nq]} (N^2 t) \rangle = u(q,t), \quad  \lim_{N \to \infty} \langle {\mc E}_{[Nq]} (N^2 t) \rangle = {\varepsilon} (q,t) \, ,
\end{equation}
where $u, \varepsilon$ are solutions of the following macroscopic diffusion equation
\begin{equation}
\label{eq:he}
\begin{cases}
\partial_t u = \gamma^{-1}\,  \partial_q^2 \, u\\
\partial_t {\varepsilon}  = (2 \gamma)^{-1} \, \partial_q^2 \, ( \varepsilon + u^2 /2)
\end{cases}
\end{equation}
with the initial conditions $u(q, 0) =u_0 (q)$, $\varepsilon (q, 0) = {\varepsilon} _0 (q)$. Eq (\ref{eq:he}) is to be solved subject to the boundary conditions
\begin{equation}
\label{eq:eqhlboundaries}
\begin{split}
&\partial_q u \, (0,t)=\partial_q u \, (1,t)=0,\\
&{\left(\ve -\cfrac{u^2}{2}\right) (0,t) = T_\ell, \quad \left(\ve -\cfrac{u^2}{2}\right) (1,t)= T_r\, .}
\end{split}
\end{equation}
The boundary condition on $\partial_q u$ comes from the fact that the length of the chain remains fixed. 

Taking the limit $t \to \infty$ in these equations we obtain the typical macroscopic profiles of the system in the NESS, i.e., a flat deformation profile $u=0$ and a linear profile ${\bar T}$ interpolating between $T_\ell$ and $T_r$,
\begin{equation}
\label{eq:bT}
\ve(q)= {\bar T} (q) =T_\ell + (T_r -T_\ell) q \, ,
\end{equation}
for the energy profile.

\subsection{The pinned chain}
Assume that the system is initially distributed according to a Gibbs local equilibrium measure associated to the energy profile ${\varepsilon}_0 (q), \; q \in [0,1]$, and define $\varepsilon (q,t)$ as the evolved profile in the diffusive scale, i.e.,
\begin{equation*}
{\ve } (q,t) = \lim_{N \to \infty} \langle {\mc E}_{[Nq]} (t N^2) \rangle \, .
\end{equation*}
Then $\varepsilon$ is the solution of the following heat equation
\begin{equation}
\begin{cases}
\partial_t \varepsilon = \partial_q (\kappa  \partial_q {\varepsilon})\, ,\\
\varepsilon (q, 0) = \varepsilon_0 (q)\, ,\\
\varepsilon (0,t)= T_\ell, \; \varepsilon (1,t) =T_r \, .
\label{eq:PHd}
\end{cases}
\end{equation}
The conductivity $\kappa$, which is independent of the temperature, is given by (\ref{eq:K}).

When $t$ goes to infinity ${\varepsilon} (q,t)$ converges to the linear profile ${{\bar T}} (q)$ (given in (\ref{eq:bT})) both for the velocity-flip and the self-consistent model. We note finally that, since the self-consistent model does not conserve energy in the bulk, we do not expect any autonomous macroscopic equations in that model.

\section{Energy Fluctuations}
\label{sec:cor}
\subsection{The pinned chain}

Our goal is to estimate the probability that in the stationary state the
empirical energy profile, $\theta^N (q)$, defined by looking at the
microscopic energy ${\mc E}_x$, for $x$ equal to the integer part of $Nq$, is close to a prescribed macroscopic energy profile $e (q)$ different from ${{\bar T}} (q)$, i.e., we want to find the large deviation function (LDF) for the NESS.

At equilibrium $T_\ell = T_r=T=\beta^{-1}$ the stationary state $\nesssc{\cdot}$ coincides with the usual Gibbs equilibrium measure $\mu^{N, {\rm eq}}_T$ and by the usual large deviations theory (see e.g.\ \cite{DZ}) we have that for any given macroscopic energy profile $e (\cdot)$
\begin{equation}
\label{eq:ld1}
{\mu}^{N,{\rm eq}}_T \left(  \theta^{N} (q) \sim e (q) \right)  \sim e^{-N V_{\rm{eq}} (e)}
\end{equation}
where the large deviation function (LDF) 
\begin{equation*}
V_{\rm{eq}} (e) =\int_0^1 \left[ \cfrac{e(q)}{T} -1- \log\left(\cfrac{e(q)}{T}\right) \right] dq 
\end{equation*}
coincides with the difference between the free energy of the system in local
thermal equilibrium (LTE) and the true equilibrium free energy with $e(q)=T$.

Out of equilibrium ($T_\ell \ne T_r$) there is also a large deviation principle \cite{Bjsp}
\begin{equation*}
\mu_{\rm{s,v}}^N \left( \theta^{N} (\cdot) \sim e (\cdot) \right)  \sim e^{-N V(e)}
\end{equation*}
but the explicit form of $V$ is in general unknown.  What is true however, is that  V depends only on two macroscopic quantities: the heat conductivity and the mobility \cite{Bjsp}. 

For the system we are interested in, the conductivity $\kappa (T)$ is given by (\ref{eq:K}) and is independent of $T$. By the Einstein relation the mobility $\chi (T)$ is equal to $\chi (T)= \kappa (T) \sigma (T)$ where $\sigma (T)$ is the static compressibility defined by the equilibrium correlation
\begin{equation}
\sigma (T) =  \sum_{x \in \ZZ}  \langle  ({\mc E}_0 -T) ({\mc E}_x -T)  \rangle^{({\rm eq},T)}.
\end{equation}
A simple computation shows that for our system  $\sigma (T) =T^2$ and Theorem 6.5 of \cite{Bjsp} applies.
It follows that $V(\cdot)$ is given by
\begin{equation}
V(e)=\int_0^1 dq \left[ \cfrac{e(q)}{F(q)} -1 -\log\left( \cfrac{e(q)}{F(q)}\right) - \log \left( \cfrac{F' (q)}{T_r -T_\ell }\right) \right]\, ,
\end{equation}
where $F$ is the unique increasing solution of
\begin{equation}
\begin{cases}
\cfrac{\partial^2_q F}{(\partial_q F)^2 } = \cfrac{F-e}{F^2}\, ,\\
F(0)=T_\ell , \; F(1) =T_r\, .
\end{cases}
\end{equation}
Surprisingly, the function $V$ is independent of the pinning value $\nu^2$ and of the intensity of the noise $\gamma$. In fact, it coincides with the LDF of the Kipnis-Marchioro-Presutti (KMP) model considered in \cite{BGL}. In that model the dynamics are entirely stochastic. 

It is now easy to derive the Gaussian fluctuations of the empirical energy. We consider a small perturbation, $e={{\bar T}} + \delta h$, of the stationary profile ${\bar e}$. The functional $V$ has a minimum at ${{\bar T}}$ so that
\begin{equation*}
V(e) = V({{\bar T}}) + \cfrac{1}{2} \, \delta^2 \, \langle h, C^{-1} h \rangle +o (\delta^2)
\end{equation*}
The operator $C$ is the covariance for the Gaussian fluctuations of the empirical energy under the invariant measure $\mu_{\rm{s,v}}^N$. The computations are the same as in \cite{BGL} and we get
\begin{equation}
C= {{\bar T}}^2 {\bf 1} + (T_r -T_\ell)^2 (-\Delta_0)^{-1}
\label{eq:Corr}
\end{equation}
where $\Delta_0$ denotes the Laplacian with Dirichlet boundary conditions on $[0,1]$.

\subsection{Unpinned Chain}
For the unpinned case, we are not able to obtain the expression of the LDF of the two conserved quantities nor for the energy alone. We conjecture that the LDF for the energy is the same for the pinned and unpinned case. Moreover, we are able (under suitable assumptions) to show that, in the unpinned case, (\ref{eq:Corr}) is still valid \cite{BKLL}.

\subsection{Self-Consistent Model} 

We show in \cite{BKLL} that the energy correlations in the NESS of a self-consistent chain are not long range in the sense that the variance of the total energy divided by $N$ is given by the LTE measure. The remainder is (at least) of order $N^{-1/2}$ in the pinned case and $N^{-1/4} \ln^2 N$ in the unpinned case.  Energy fluctuations are thus dominated by the local equilibrium term. Moreover, in the pinned chain the NESS is reached exponentially fast on a microscopic time-scale. Therefore, the energy profile is then given by the linear equilibrium profile for all macroscopic times.

\section{Numerical simulations}
\label{sec:num}

We have performed numerical simulations  in
the velocity-flip model, both with and without bulk pinning.  In these simulations, the initial position and momentum of each particle was chosen randomly, from a centered uniform distribution in the intervals[$-\sqrt{2T/k},\sqrt{2T/k}$] and [$-\sqrt{2 T m},\sqrt{2  T m}$] where $T= \frac{T_l +T_r}{2}$ and $k$ is the interparticle harmonic potential, and we have set $m=1$ and $k_B=1$ (the mass and Boltzmann's constant respectively).  The numerical solution to the  dynamics was carried out by choosing a small step $\delta t$, and at each step updating the position and momentum of the particles first according to the Hamiltonian evolution, then for the heat baths, and finally taking into account the velocity flips.  The solution to the Hamiltonian part was obtained by
performing an Euler integration, implemented through the velocity-Verlet algorithm.
A numerical approximation was used to determine the white-noise forces for the Langevin heat baths, and a standard scheme was implemented to generate the Poisson processes for the velocity flips at every site.  The state of the system was judged to be sufficiently close to the steady state when the currents $\mean{j^e_x}$ were essentially constant throughout the chain.

The total energy $\mathcal{H}$ is given by (\ref{eq:energy}), and our goal is to estimate numerically its total fluctuations, by measuring the observable
$$s_N = N \cfrac{\nessvf{{\mc H^2}} - \nessvf{{\mc H}}^2}{(\nessvf{\mc H})^2}\, .$$
Using equation (\ref{eq:Corr}) the measured fluctuations in the total energy should be given by
\begin{equation*}
s_{\infty} = \lim_{N \to \infty} s_N = s_{\infty}^{\rm{loc. eq.}} + c_{\infty}\, ,
\end{equation*}
where $s_{\infty}^{\rm{loc. eq.}} = {\int_0^1 {\bar T}^2 (q) dq}/{\left[ \int_0^1 {\bar T} (q) dq\right]^2}$ is the value we would obtain by a local equilibrium approximation, and $c_{\infty}= \frac{1}{12} {(T_r -T_\ell)^2} /{\left[ \int_0^1 {\bar T} (q) dq\right]^2}$ is the correction due to the long-range correlations.  The prefactor $\frac{1}{12}$ is obtained by integrating over $((-\Delta_0)^{-1} 1) (q)= q(1-q)/2$.  Computing the remaining explicit integrals yields then
\begin{equation*}
s_{\infty}=\cfrac{4 T_\ell T_r + \frac{5}{3} (T_r- T_\ell)^2}{(T_\ell + T_r)^2} \, .
\end{equation*}

\begin{table}
\scalebox{0.8}{
\begin{tabular}{|p{1cm}|p{2cm}|p{1cm}|p{2cm}|p{1.5cm}|p{1.5 cm}| p{1.5cm}|}
\hline
$N$ & $T_\ell$, $T_r$ & $\gamma$ & ${{\tilde s}_N}$ & Error $| s_{\infty} -{\tilde s}_N |$ & $s_{\infty}^{\rm{loc. eq.}}$ & $s_{\infty}$\\
\hline
100 & 8,1  & 0.1 & 1.40 &0.01 & \multirow{3}{*}{1.20} & \multirow{3}{*}{1.40}\\
200 & 8,1 & 0.1 & 1.39  &0.01 & & \\
400 & 8,1 & 0.1 & 1.42 &0.02 &  &\\
\hline
\end{tabular}
}
\caption{Total energy variation in the unpinned model ($\nu=0$).\label{table:unpinned}}
\end{table}

\begin{table}
\scalebox{0.8}{
\begin{tabular}{|p{1.5cm}|p{1.5cm}|p{1cm}|p{1cm}|p{1.5cm}|p{1.5cm}|p{1.5cm}| p{1.5cm}| }
\hline
N&$T_\ell, T_r$ & $\nu^2$ & $\gamma$ & ${\tilde s}_N$ & Error $|s_{\infty} -{\tilde s}_N|$& $s_{\infty}^{\rm{loc. eq.}}$     & $ s_{\infty} $\\
\hline
  200&8,1  & 0.25 & 0.1 & 1.38 &0.01 & \multirow{5}{*}{1.20} & \multirow{5}{*}{1.40}\\
 200&8,1  & 0.5 & 0.1& 1.39  &0.01  &  & \\
 200  &8,1& 0.25 &1.0& 1.39 &0.02   &   &\\
 400 &8,1& 0.25 & 0.1 &1.39 & 0.01  &  &\\
 800  &8,1& 0.25&0.1&1.46&0.05& &\\
\hline
200 &5,1& 0.25 & 0.1 &1.30 & 0.01 & 1.15& 1.30\\
\hline
\end{tabular}
}

\caption{Total energy variation in the pinned model ($\nu>0$).\label{table:pinned}}

\end{table}

To get good statistics, several realizations of the initial data were considered.
The results from numerical simulations for the variation in total energy, denoted by ${\tilde s}_N$, are collected in Tables \ref{table:unpinned} and \ref{table:pinned}. These results are compared with the theoretical estimate, $s_{\infty}$, and with $s_{\infty}^{\rm{loc. eq.}}$.  We see that there is a very close match between the predicted $s_{\infty}$ and the measured values.

\section{Higher Dimensional Systems}

\label{sec:conc}
The NESS for the self-consistent crystal with coordinates $q_{i_1, \ldots, i_D} \, , i_{\alpha}=1,2, \ldots, N$, with thermal reservoirs with temperatures $T_{\ell}$ ($T_r$) in contact with all oscillators having values $i_1=1\,  (i_1=N)$ and periodic boundary condition in the transverse directions was solved in \cite{BLL04}. As shown in \cite{DKL} this has the same covariances as the corresponding velocity flip model. In particular the heat conductivity $\kappa$ is given by 
\begin{equation}
\kappa = \frac{1}{\gamma} \int_{[0,1]^{D-1}} \frac{d^{D-1} {\bf y}}{2+\tilde{\nu}({\bf y})^2 + \sqrt{\tilde{\nu}({\bf y})^2 (4+ \tilde{\nu}({\bf y})^2)}}
\label{eq:HDk}
\end{equation}
where $\tilde{\nu}({\bf y})^2 = \nu^2 + 2 \sum_{i=1}^{D-1} (1-\text{cos}(2\pi y_i))$. The same argument which gives $\musv$ as a superposition of Gaussians in 1D also holds in $D > 1$.  The hydrodynamical equation for the pinned velocity flip case will again be that of the form (\ref{eq:PHd}) 
\begin{equation*}
\frac{\partial}{\partial t} \varepsilon({\bf q},t) = \kappa \sum_{\alpha=1}^{d} \frac{\partial^2}{\partial q_{\alpha}^2} \varepsilon
\end{equation*} 
where $\kappa$ is given by (\ref{eq:HDk}). Unfortunately we do not have any information about the LDF in the $D>1$ case.

\bigskip

\noindent
{\textsc{Acknowledgements} We thank S.~Olla for useful discussions. C.~Bernardin acknowledges the support of the French Ministry of Education through the ANR-10-BLAN 0108 grant.  J.~Lukkarinen was supported by the Academy of Finland. The authors thank the Rutgers University, the Fields Institute, the IHES and the IHP for their hospitality. This work was supported in part by NSF Grants DMR 08-02120 and by AFOSR Grant FA9550-10-1-0131.}

\end{document}